\def\draftversion{false}

\RequirePackage{ifthen}
\ifthenelse{\equal{\draftversion}{true}}{
  \documentclass[aps,prb,10pt,galley,amsmath,amssymb,showpacs]{revtex4}
}{
  \documentclass[aps,prb,10pt,twocolumn,amsmath,amssymb,showpacs]{revtex4-1}
}

\usepackage{graphicx}
\usepackage[usenames]{color} 
\usepackage{bm} 
\usepackage{dcolumn} 
\usepackage{ulem} 

\ifthenelse{\equal{\draftversion}{true}}{
  \marginparwidth 2.7in
  \marginparsep 0.5in
  \newcounter{comm} 
  \def\commnext{\stepcounter{comm}}
  \def\commtext{{\bf\color{blue}[\arabic{comm}]}}
  \def\commmar{{\bf\color{blue}[\arabic{comm}]}}
  \def\dvm#1{\commnext\marginpar{\small DV\commmar: #1}\commtext}
  \def\mym#1{\commnext\marginpar{\small MY\commmar: #1}\commtext}
  \def\wwm#1{\commnext\marginpar{\small Weida\commmar: #1}\commtext}
  \def\mlab#1{\marginpar{\small\bf #1}}
  
}{
  \def\dvm#1{}
  \def\mym#1{}
  \def\wwm#1{}
  \def\mlab#1{}
  
}

\def\beq{\begin{equation}}
\def\eeq{\end{equation}}

\def\CrO{Cr$_2$O$_3$}
\def\HoMO{HoMnO$_3$}
\def\ErMO{ErMnO$_3$}
\def\YbMO{YbMnO$_3$}
\def\LuMO{LuMnO$_3$}
\def\LuFO{LuFeO$_3$}
\def\RMO{$R$MnO$_3$}
\def\RFO{$R$FeO$_3$}
\def\Zm{Z^{\rm m}}
\def\Ze{Z^{\rm e}}
\def\Op{O$_{\rm P}$}
\def\Ot{O$_{\rm T}$}

\begin{document}


\title{Magnetic charges and magnetoelectricity in hexagonal rare-earth
   manganites and ferrites }

\author{Meng Ye}
\email{mengye@physics.rutgers.edu}
\affiliation{Department of Physics \& Astronomy, Rutgers University,
Piscataway, New Jersey 08854, USA}

\author{David Vanderbilt}
\affiliation{Department of Physics \& Astronomy, Rutgers University,
Piscataway, New Jersey 08854, USA}

\date{\today}

\begin{abstract}

Magnetoelectric (ME) materials are of fundamental interest and
show broad potential for technological applications.
Commonly the dominant contribution to the ME response is the
lattice-mediated one, which is proportional to both the Born electric
charge $\Ze$ and its analogue, the dynamical magnetic charge $\Zm$.
Our previous study has shown that exchange striction acting on
noncollinear spins induces much larger magnetic charges than those
that depend on spin-orbit coupling. The hexagonal manganites \RMO\ and
ferrites \RFO\ ($R$ = Sc, Y, In, Ho-Lu) exhibit strong couplings between
electric, magnetic and structural degrees of freedom, with the
transition-metal ions in the basal plane antiferromagnetically
coupled through super-exchange so as to form a 120$^\circ$ noncollinear
spin arrangement. Here we present a theoretical study of the magnetic
charges, and of the spin-lattice and spin-electronic ME constants, in
these hexagonal manganites and ferrites, clarifying the conditions
under which exchange striction leads to an enhanced $\Zm$ values
and anomalously large in-plane spin-lattice ME effects.


\end{abstract}

\maketitle


\section{Introduction}
The cross-coupling between magnetic, electric, and elastic properties
can lead to a plethora of novel and profound physical phenomena, with
potentially broad and innovative applications. Magnetoelectric (ME)
effects are those in which the electric polarization $\mathbf P$
responds to an applied magnetic field $\mathbf H$, or magnetization
$\mathbf M$ responds to an applied electric field $\bm{\mathcal E}$.
The ME coupling (MEC) between magnetic and electric properties has
motivated intense experimental and theoretical investigations in bulk
single crystals, thin films, composite layers, and organic-inorganic
hybrid materials in recent years. \cite{Fiebig05,Spaldin05,Scott06,
Tokura07,Cheong&Mostovoy07,Ramesh&Spaldin07,Wang09,Fiebig&Spaldin09,
Rivera09,Khomskii09,Birol12}

At the linear-response level, the
linear MEC tensor $\alpha$ is defined as
\beq
\alpha_{\beta\nu}=
\frac{\partial P_\beta}{\partial H_\nu}\Big\vert_{\bm{\mathcal{E}}}
=\mu_0\frac{\partial M_\nu}{\partial{\mathcal{E}}_\beta}\Big\vert_{\mathbf H}\,,
\label{eq:ME}
\eeq
where indices $\beta$ and $\nu$ denote the Cartesian directions and
$\mu_0$ is the vacuum permeability. From a theoretical point of view,
the linear ME effect can be decomposed into electronic (frozen-ion),
ionic (lattice-mediated), and strain-mediated responses. \cite{Birol12}
Each term can be further subdivided into spin and orbital contributions
based on the origin of the induced magnetization.  As the orbital
moment is usually strongly quenched on the transition-metal sites,
most phenomenological and first-principles studies have focused
on the spin-electronic \cite{Bousquet11} and the spin-lattice
\cite{Iniguez08,KITPite09,Das14} contributions. The lattice response
can be written, following Ref.~\onlinecite{Iniguez08}, as
\beq
\alpha^{\rm {latt}}_{\beta\nu} =
\Omega_0^{-1} \mu_0 \Ze_{m\beta} \, (K^{-1})_{mn} \, \Zm_{n\nu} \,,
\label{eq:MEspin-lat}
\eeq
(sum over repeated indices implied),
i.e., as a matrix product of the dynamical Born electric charge $\Ze$,
the inverse force-constant matrix $K^{-1}$, and the dynamical magnetic
charge $\Zm$, where $m$ and $n$ are composite labels for an atom and
its displacement direction. $\Omega_0$ is the unit cell volume. Note
that $\Zm$ is the magnetic analog of the dynamical Born charge, and is
defined as
\beq
\Zm_{m\nu}=\Omega_0\frac{\partial M_\nu}{\partial u_m}\Big\vert_{\bm{\mathcal E},\mathbf H,\mathbf \eta}
=\mu_0^{-1}\frac{\partial F_m}{\partial H_\nu}\Big\vert_{\bm{\mathcal E},\mathbf{u},\mathbf\eta} \,,
\label{eq:Zm}
\eeq
where $u_m$ is an internal displacement, $F_m$ is an atomic force, and
$\mathbf\eta$ is a homogeneous strain. In principle, $\Zm$ has both
spin and orbital parts, corresponding respectively to spin and orbital
contributions to $M_\nu$, or Zeeman and ${\bf p}\cdot{\bf A}$ terms
induced by $H_\nu$, but we shall focus on the spin part in the following.
Our previous first-principles study has shown that exchange striction
acting on noncollinear spin structures induces much larger magnetic
charges than when $\Zm$ is driven only by spin-orbit coupling (SOC).
Therefore, exchange striction provides a promising mechanism for
obtaining large MECs. \cite{Ye14}

The hexagonal manganites \RMO\ and ferrites \RFO\ ($R$ = Sc, Y, In, and
Ho-Lu) form an interesting class of materials exhibiting strong couplings
between electric, magnetic, and structural degrees of freedom. \cite{Mostovoy10}
A series of first-principles \cite{VanAken04,Fennie05,Kumagai13,Das14}
and phenomenological \cite{Artyukhin13} studies have greatly enhanced
our understanding of the coupled properties.
The ferroelectricity is induced by the structural trimerization, and
the direction of the spontaneous polarization is related to the
trimerization pattern. \cite{Fennie05} An interesting ``cloverleaf''
pattern formed from interlocking domain walls between structural and
ferroelectric domains has been found in hexagonal \RMO\ and is now
understood in terms of Landau theory. \cite{Choi10, Artyukhin13,Lin14}
Hexagonal \RMO\ and \RFO\ have rich magnetic phase diagrams and show
considerable potential for manipulation and practical applications.
\cite{Fiebig03,Yen07,Lorenz13} The magnetic order has two different
origins, with the transition-metal Mn$^{3+}$ or Fe$^{3+}$ sublattices
ordering first, often followed by ordering of the rare-earth ions
$R^{3+}$ at lower temperature. The magnetic anisotropy is easy-plane
and easy-axis for $3d$ and $4f$ spins
respectively; the $3d$ moments are antiferromagnetically
coupled through superexchange so as to form a 120$^\circ$ noncollinear
arrangement in the $x$-$y$ plane, while the 4$f$ rare-earth moments are
collinear along the hexagonal $z$ axis.

The low-temperature magnetic phases of \RMO\ and \RFO\ allow a linear ME
effect to be present. The recently developed ME force microscopy technique
has been used successfully to observe the ME domains in \ErMO. \cite{Geng14}
In that work, a large ME component $\alpha_{zz}$ $\sim$\,13 ps/m has been
measured at 4\,K, which is below the Mn$^{3+}$ ordering temperature of
81\,K but above the Er$^{3+}$ ordering temperature of 2\,K.
However, first-principles calculations predict that the SOC-induced
spin-lattice $\alpha_{zz}$ arising from the Mn$^{3+}$ order is
0.7-1.0\,ps/m. \cite{Das14} This discrepancy suggests that the dominant
ME effect in the hexagonal $\hat{z}$ direction is mediated by the
Er$^{3+}$ 4$f$ electrons in \ErMO. The in-plane ME effect, which has not
been measured  or calculated, has a different origin. It is dominated by
an exchange-striction mechanism on the Mn$^{3+}$ sublattice, because the
noncollinear spin pattern is sensitive to the lattice distortion.
Thus, hexagonal \RMO\ and \RFO\ are good candidates to show
exchange-striction enhanced magnetic charges and anomalously
large spin-lattice MECs.

In this work, we use first-principles density-functional methods to
study the magnetic charges and the spin-induced MECs arising from the
$3d$ electrons in hexagonal \HoMO, \ErMO, \YbMO, \LuMO, and \LuFO. For
the transverse magnetic charge components and MECs, we also provide a
comparison between results induced solely by exchange striction and ones
including SOC. Our results confirm that the exchange striction greatly
enhances the in-plane magnetic charges, while the SOC contribution is
minor for most components except on Mn atoms.
However, the effect of SOC on the MECs is surprisingly large in
many cases. This occurs because the exchange-striction contribution
tends to be reduced by cancellations between modes, while the SOC
contribution is mainly associated with a few low-frequency modes.
The in-plane ME responses are discussed case
by case, and the conditions under which exchange striction leads to
anomalously large in-plane spin-lattice MECs are clarified.

The paper is organized as follows. In Sec.~\ref{sec:RMOstru} and
\ref{sec:RFOstru} we introduce the geometric structure and
magnetic order of hexagonal
\RMO\ and \RFO. In Sec.~\ref{sec:sym} we analyze the tensor symmetries
of the Born charges, magnetic charges and MECs in two different
magnetic phases of \RMO\ and \RFO. The computational details are
described in Sec.~\ref{sec:DFT}. The results and discussion of Born
charges, magnetic charges and MECs in \RMO\ and \LuFO\ are presented in
Sec.~\ref{sec:results}.
We summarize and give our conclusions in Sec.\ref{sec:summary}.

\section{PRELIMINARIES}
\subsection{Hexagonal \RMO} \label{sec:RMOstru}
\begin{figure}
  \includegraphics[width=8.5cm]{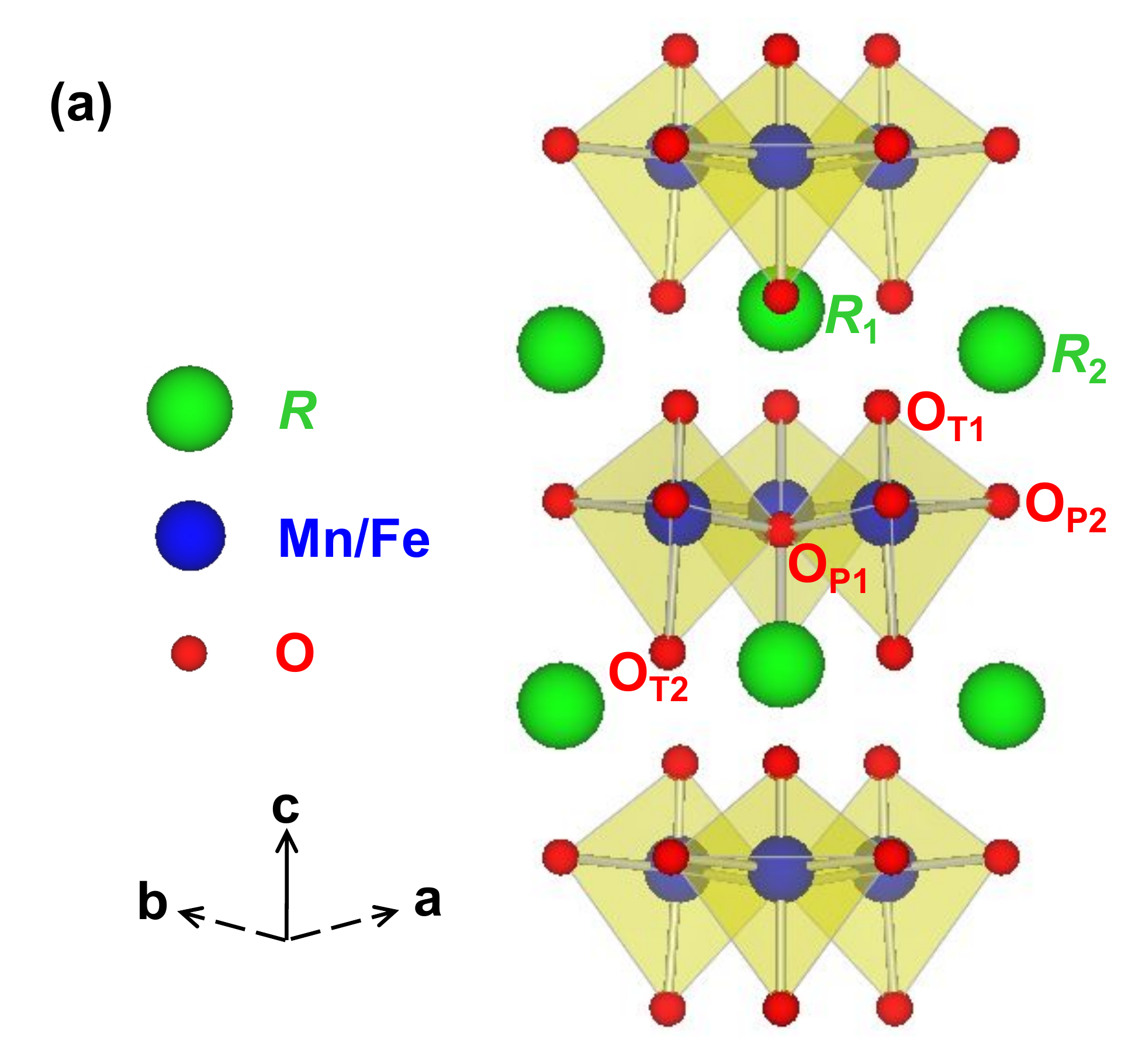}
  \includegraphics[width=8.5cm]{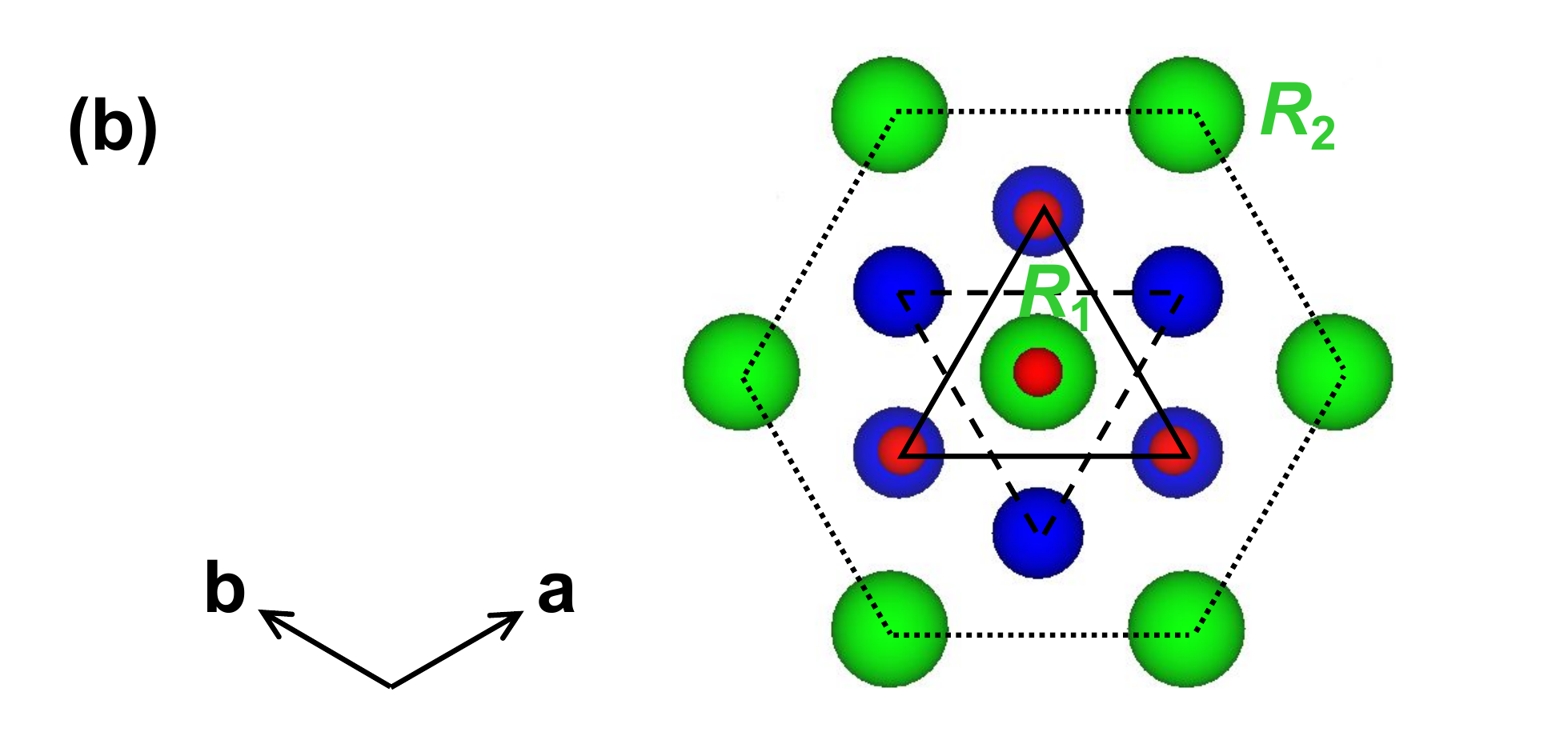}
  \caption{Structure of ferroelectric hexagonal
  \RMO\ (6 f.u.\ per primitive cell).
  (a) Side view from [110]. (b) Plan view from [001]; dashed
  (solid) triangle indicates three Mn$^{3+}$
  connected via Op$_1$ to form a triangular sublattice at $z=0$
  ($z=1/2$).}
  \label{fig:RMOstru}
\end{figure}

Above the structural transition temperature $T_{\rm c}$ $\sim$\,900\,-
\,1500\,K, the hexagonal manganites \RMO\ ($R$ = Sc, Y, In, and Ho-Lu)
are paraelectric insulators. The space group is P$6_3$/mmc with two
formula units (f.u.) per primitive cell. Below $T_c$, the size mismatch
between the small-radius $R^{3+}$ ion and the large MnO$_5$ bipyramid
leads to an inward tilting of the three corner-shared MnO$_5$ polyhedra
and an associated ``one-up/two-down'' buckling of the $R^{3+}$ ion
layer, as shown in Fig.~\ref{fig:RMOstru}. The transition triples
(``trimerizes'') the unit cell, lowers the structural symmetry to
P$6_3$cm, and induces ferroelectricity. As the induced polarization is
nonlinearly coupled to the trimerization, these systems are
improper ferroelectrics.\cite{VanAken04,Fennie05,Artyukhin13}

\begin{figure}
  \includegraphics[height=2.7cm]{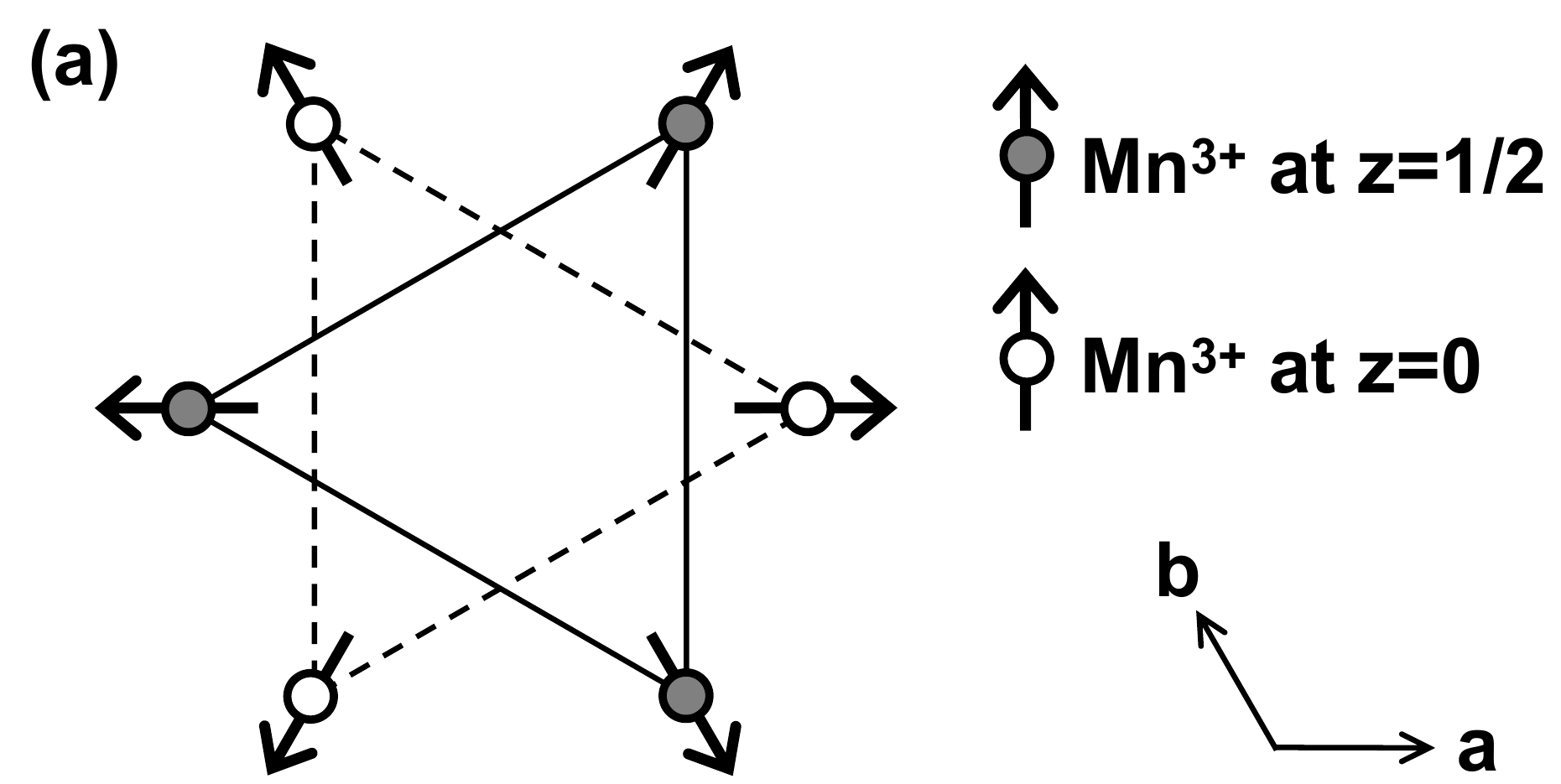}
  \includegraphics[height=2.7cm]{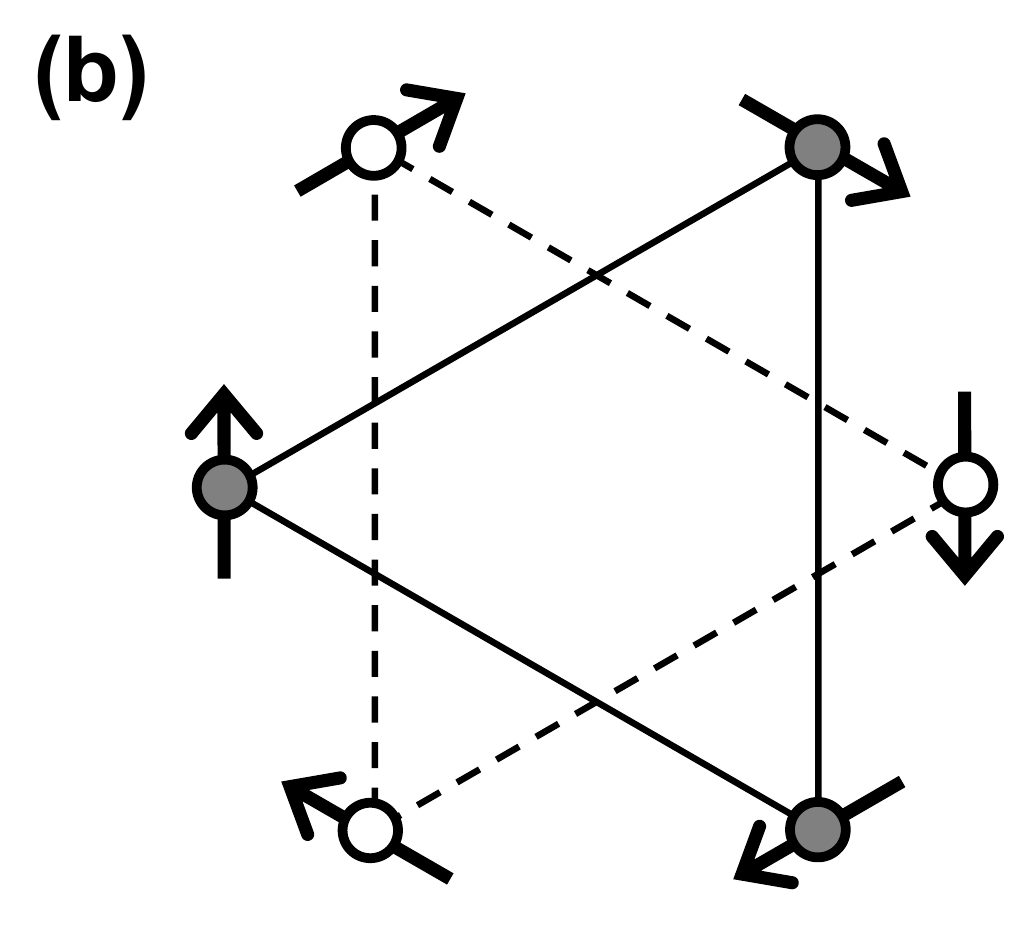}
  \caption{Magnetic phases of hexagonal \RMO\ and \RFO.
  Mn$^{3+}$ ions form triangular sublattices at $z=0$ (dash line)
  and $z=1/2$ (solid line).
  (a) A$_2$ phase with magnetic symmetry P6$_3$c$'$m$'$;
  spins on a given Mn$^{3+}$ layer point all in or all out.
  (b) A$_1$ phase with the magnetic symmetry P6$_3$cm,
  with Mn$^{3+}$ spins pointing tangentially to form
  a vortex pattern. The A$_1$ and A$_2$ phases
  differ by a 90$^\circ$ global rotation of the spins.
  The B$_1$ and B$_2$ phases can be obtained from A$_2$ and A$_1$
  by reversing the spins on the dashed triangles.}
  \label{fig:Aphase}
\end{figure}

The Mn$^{3+}$ magnetic order develops below the N\'{e}el temperature
$T_{\rm N}$ of $\sim$\,70\,-\,130\,K. The in-plane Mn-O-Mn superexchange
determines the noncollinear 120$^\circ$ antiferromagnetic (AFM)
order on the Mn$^{3+}$ triangular lattice. On the other hand,
the inter-plane Mn-O-$R$-O-Mn exchange, which is two orders of magnitude
weaker than the in-plane exchange, modulates the relative spin
directions between two consecutive Mn planes. \cite{Fiebig03,Das14}
At temperatures lower than $\sim$\,5.5\,K, the rare-earth ions with
partially filled 4$f$ shells develop collinear spin order along the
hexagonal $z$ direction. For the Mn$^{3+}$ order, there are four
distinct magnetic phases, namely A$_1$ (P6$_3$cm), A$_2$ (P6$_3$c$'$m$'$),
B$_1$ (P6$_3'$cm$'$), and B$_2$ (P6$_3'$c$'$m). The linear ME effect
exists only in A$_1$ and A$_2$ phases. The A$_1$ and A$_2$ phases are
shown in Fig.~\ref{fig:Aphase}; the B$_1$ and B$_2$ phases can be
obtained from A$_2$ and A$_1$ by reversing the spins on the dashed
triangles.  From previous experiments, it is known that at zero
temperature without a magnetic field, \HoMO\ is in the A$_1$ phase, while
\ErMO, \YbMO, and \LuMO\ are not in either A phase.  Under a weak
magnetic field along the $\hat{z}$ direction, \ErMO\ and \YbMO\
undergo a transition into the A$_2$ phase.\cite{Fiebig03,Yen07,Lorenz13}

\subsection{Hexagonal \RFO} \label{sec:RFOstru}
Epitaxially grown thin-film hexagonal \RFO\ has a similar structure
as hexagonal \RMO, with improper ferroelectricity below $\sim$\,1000\,K.
Replacing Mn$^{3+}$ with Fe$^{3+}$ introduces larger spin moments and
stronger super-exchange interactions in the basal plane. In a recent
experiment, AFM order has been found to develop at $T_{\rm N}=440$\,K
followed by a spin-reorientation transition below $T_{\rm R}=130$\,K
in \LuFO.\cite{Wang13} It has also been confirmed that below 5\,K, the
magnetic structure of \LuFO\ is that of the A$_2$ phase.\cite{Disseler14}

\subsection{Symmetry} \label{sec:sym}

 Our purpose is to understand the mechanisms that generate large
magnetic charges that may in turn induce anomalously large spin-lattice
MECs. Therefore, we focus on the A$_1$ and A$_2$ magnetic phases,
shown in Fig.~\ref{fig:Aphase}, which allow a linear MEC to exist.
 \ErMO, \YbMO, and \LuMO\ actually adopt other phases as their
ground-state magnetic order at low temperature. Nevertheless, we
include them for purposes of comparison when calculating the
properties of the hexagonal \RMO\ materials in the A$_2$ phase.
We also study \LuFO\ in the A$_2$ phase, and for \HoMO\ we study both
the A$_1$ and A$_2$ phases.

\begin{figure}
  \includegraphics[width=8cm]{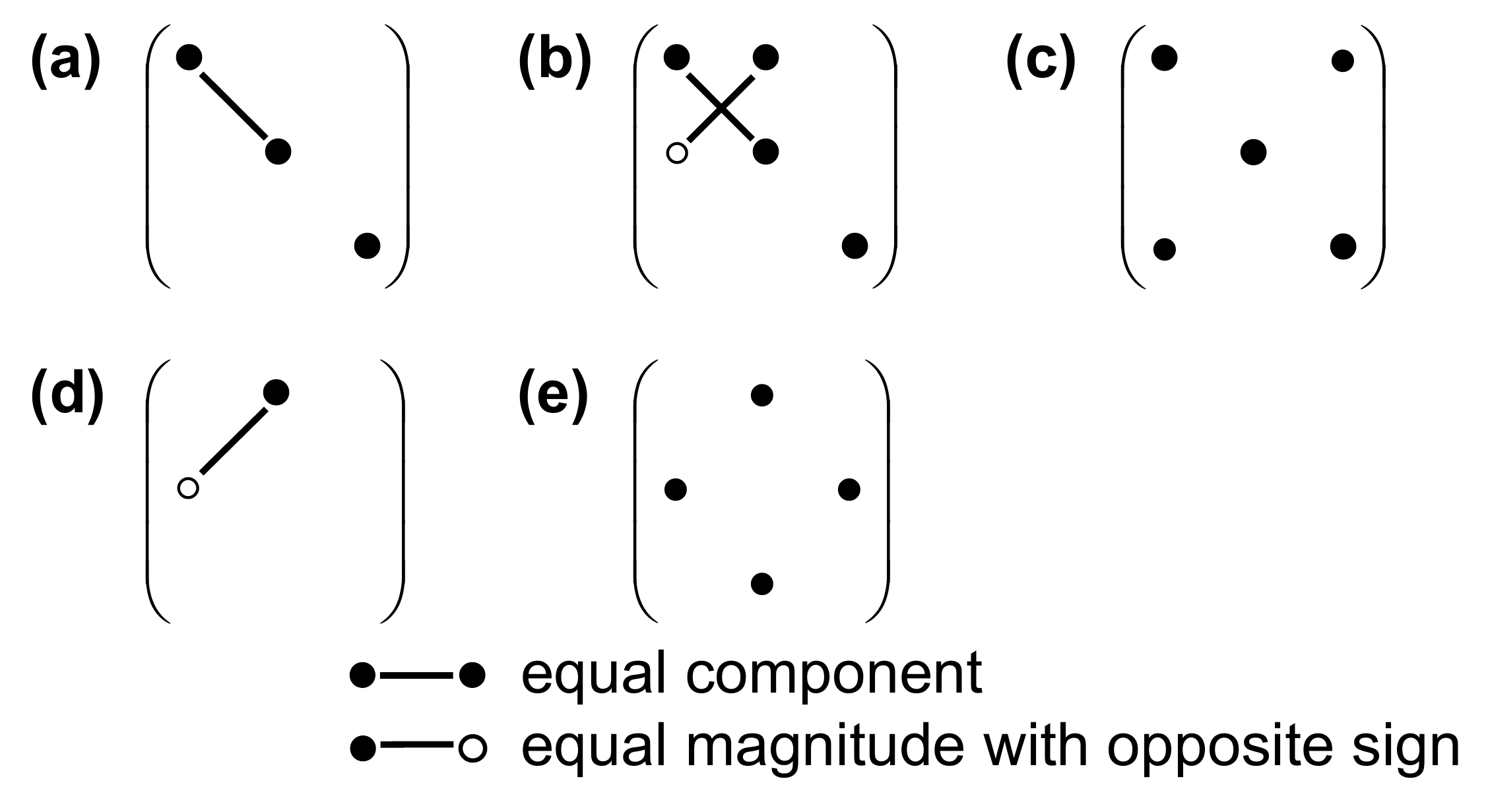}
  \caption{Symmetry patterns of Born charges, magnetic charges and ME
  tensors in \RMO\ and \RFO.
  (a) Tensor form of the ME coupling in the A$_2$ phase,
  Born charges on $R_1$ and \Op$_1$ sites in either A phase,
  and magnetic charges on the same sites in the A$_2$ phase.
  (b) Tensor form of the Born and magnetic charges on $R_2$ and \Op$_2$
  sites in either A phase.
  (c) Tensor form of the Born charges on Mn, Fe, \Ot$_1$, and \Ot$_2$
  sites lying on an $M_y$ mirror plane in either A phase, and of
  the magnetic charges on the same sites in the A$_2$ phase.
  (d) Tensor form of the ME coupling in the A$_1$ phase, and of
  the magnetic charges on R$_1$ and \Op$_1$ sites in the A$_1$
  phase.
  (e) Tensor form of the magnetic charges on Mn, Fe, \Ot$_1$, and
  \Ot$_2$ sites lying on an $M_y$ mirror plane in the A$_1$ phase.}
  \label{fig:tensorsym}
\end{figure}

 The A$_1$ and A$_2$ phases have the same P6$_3$cm structural symmetry,
so the forms of the atomic Born charge tensors in the two phases are the same.
The Born charges for $R_1$ and \Op$_1$ take the tensor form shown in
Fig.~\ref{fig:tensorsym}(a), while those of $R_2$ and \Op$_2$ have
the symmetry pattern shown in Fig.~\ref{fig:tensorsym}(b).
For the Mn, Fe, \Ot$_1$, and \Ot$_2$ sites lying on a vertical
$M_y$ mirror plane, the Born charges are as given in
Fig.~\ref{fig:tensorsym}(c); for the partner sites related by
rotational symmetry, the tensors also need to be rotated accordingly.

The symmetry forms of the atomic magnetic charge tensors can be derived
from the on-site magnetic point symmetries. For the A$_1$ phase,
the magnetic space group is P6$_3$cm
and the magnetic charges of $R_1$ and \Op$_1$ take the forms given in
Fig.~\ref{fig:tensorsym}(d);
those for $R_2$ and \Op$_2$ have the tensor symmetry
shown in Fig.~\ref{fig:tensorsym}(b); and for Mn, Fe, \Ot$_1$, and \Ot$_2$
they can be written in the form of Fig.~\ref{fig:tensorsym}(e). For the A$_2$
phase, the magnetic group is P6$_3$c$'$m$'$; all the
improper operators are associated with the time-reversal operation, so
the magnetic charges have the same tensor forms as the Born charges.

A symmetry analysis of the structure and the magnetic space group
identifies the phonon modes that couple to the electromagnetic field.
The infrared (IR)-active phonon modes that couple to the electric
field are the longitudinal $A_1$ modes and the transverse $E_1$ modes,
\beq
\Gamma_{\rm {IR}}=10A_1+15E_1\,,
\eeq
including the three acoustic modes.  The magnetization is generated by
phonon modes that couple to the magnetic field. In the A$_1$ phase, the
magneto-active phonon modes are the longitudinal $A_2$ modes and the
transverse $E_1$ modes,
\beq
\Gamma^{\rm{A}_1}_{\rm {mag}}=5A_2 + 15E_1\,,
\eeq
where one pair of acoustic $E_1$ modes are included.
In the A$_2$ phase, on the other hand, the IR- and magneto-active
phonon modes are identical, since the magnetic and Born charge tensors
have the same form in this case.

For the MECs in the A$_1$ phase, as the longitudinal IR-active and
magneto-active modes are mutually exclusive, the ME tensor takes the form
of Fig.~\ref{fig:tensorsym}(d), which does not have a longitudinal ME
component. For the A$_2$ magnetic phase, the $A_1$ and $E_1$ modes are
both IR-active and magneto-active, so that the ME tensor has both
longitudinal and transverse components and adopts the form shown in
Fig.~\ref{fig:tensorsym}(a).

\begin{table}
\begin{center}
\begin{ruledtabular}
    \caption{Atomic Born charge tensors $\Ze$ (in units of $|e|$)
    for \LuMO\ and \LuFO\ in the A$_2$ phase. TM\,=\,Mn or Fe.}
    \begin{tabular}{cddccdd}
        &\multicolumn{1}{c}{\LuMO}
        &\multicolumn{1}{c}{\LuFO} &&
        &\multicolumn{1}{c}{\LuMO}
        &\multicolumn{1}{c}{\LuFO}\\ \hline
    $\Ze_{xx}$(Lu$_1$)  & 3.61& 3.79&& $\Ze_{xz}$(\Ot$_1$) & 0.19& 0.11\\
    $\Ze_{zz}$(Lu$_1$)  & 4.12& 3.94&& $\Ze_{zz}$(\Ot$_1$) &-3.19&-3.21\\
    $\Ze_{xx}$(Lu$_2$)  & 3.66& 3.84&& $\Ze_{xx}$(\Ot$_2$) &-1.90&-2.15\\
    $\Ze_{yx}$(Lu$_2$)  & 0.13& 0.15&& $\Ze_{zx}$(\Ot$_2$) &-0.20&-0.19\\
    $\Ze_{zz}$(Lu$_2$)  & 3.96& 3.88&& $\Ze_{yy}$(\Ot$_2$) &-1.85&-2.13\\
    $\Ze_{xx}$(TM)      & 3.17& 2.96&& $\Ze_{xz}$(\Ot$_2$) &-0.18&-0.11\\
    $\Ze_{zx}$(TM)      & 0.44& 0.21&& $\Ze_{zz}$(\Ot$_2$) &-3.33&-3.30\\
    $\Ze_{yy}$(TM)      & 3.26& 3.01&& $\Ze_{xx}$(\Op$_1$) &-3.00&-2.40\\
    $\Ze_{xz}$(TM)      & 0.07&-0.02&& $\Ze_{zz}$(\Op$_1$) &-1.54&-1.61\\
    $\Ze_{zz}$(TM)      & 3.95& 4.16&& $\Ze_{xx}$(\Op$_2$) &-3.05&-2.45\\
    $\Ze_{xx}$(\Ot$_1$) &-1.92&-2.19&& $\Ze_{yx}$(\Op$_2$) &-0.03&-0.02\\
    $\Ze_{zx}$(\Ot$_1$) & 0.25& 0.25&& $\Ze_{zz}$(\Op$_2$) &-1.43&-1.52\\
    $\Ze_{yy}$(\Ot$_1$) &-2.00&-2.28 \\
    \end{tabular}\label{tab:Ze}
\end{ruledtabular}
\end{center}
\end{table}

\subsection{First-principles methods} \label{sec:DFT}
Our calculations are performed with plane-wave density functional
theory (DFT) implemented in VASP \cite{VASP} using the generalized-gradient
approximation parametrized by the Perdew-Burke-Ernzerhof functional.
\cite{PBE} The ionic core environment is simulated by projector
augmented wave (PAW) pseudopotentials, \cite{PAW} and the 4$f$
electrons are placed in the PAW core. We use a Hubbard $U=4.5$\,eV
and $J=0.95$\,eV on the $d$ orbitals of the Mn/Fe atoms, and the moment
on the rare-earth ions are not considered. \cite{Das14}
The structures are fully relaxed in the DFT+U \cite{DFTU}
calculations with their non-collinear spin arrangements in two cases,
when SOC is present and when it is absent.
In our noncollinear magnetization
calculation, a high cutoff energy 700\,eV and a tight energy error
threshold $1.0\times10^{-9}$\,eV are necessary to get fully converged
magnetic properties.  The Born effective charge tensors and
the $\Gamma$-point force-constant matrices are obtained using
linear-response methods in the absence of SOC.
The dynamical magnetic charges are computed by applying a
uniform Zeeman field \cite{Bousquet11} to the crystal and computing the
resulting forces. Polarization is calculated using the Berry phase
formalism. \cite{ModernPolar} A $4\times4\times2$ $\Gamma$-centered
k-point mesh is used in the calculations.

\section{Results and discussion}
\label{sec:results}
\subsection{Born charge and force-constant matrix} \label{sec:ZeandK} 

 The $f$ electrons are not included in our calculations for the
hexagonal \RMO\ class of materials, so the major differences between
compounds result from the variation of the rare-earth radius;
the trimerization tends to increase as the radius of the
rare-earth element decreases. Because of the similarity in the
geometric structures, the dielectric and phonon properties are
almost identical in the \RMO\ compounds, regardless of the magnetic
ordering. In Tables~\ref{tab:Ze} and \ref{tab:Cn} we list the Born
charge tensors and the eigenvalues of the force-constant matrix for
the IR-active modes of \LuMO\ and \LuFO. Only small differences are
observed between \LuMO\ and \LuFO, reflecting the different
transition-metal atom. The results for the other \RMO\ compounds
are quite similar to those of \LuMO\ and are given for completeness
in the Supplement.

\begin{table}
\begin{center}
\begin{ruledtabular}
    \caption{Eigenvalues of the force-constants matrix ($\rm{eV}/{\AA}^2$)
    for IR-active modes in \LuMO\ and \LuFO\ in the A$_2$ phase,
    excluding translational modes.}
    \begin{tabular}{ddcdd}
        \multicolumn{2}{c}{$A_1$ modes} &&\multicolumn{2}{c}{$E_1$ modes} \\
        \multicolumn{1}{c}{\LuMO}  &\multicolumn{1}{c}{\LuFO}&
        &\multicolumn{1}{c}{\LuMO} &\multicolumn{1}{c}{\LuFO}\\ \hline
        4.24  &3.48  &&3.32  &3.56 \\
        7.44  &6.70  &&4.68  &4.62 \\
        8.74  &8.41  &&6.73  &6.97 \\
        11.51 &11.47 &&7.35  &8.09 \\
        14.01 &12.03 &&8.63  &8.83 \\
        15.60 &15.59 &&9.56  &9.24 \\
        22.66 &20.53 &&11.36 &11.37\\
        25.87 &22.83 &&12.46 &12.46\\
        35.82 &28.46 &&13.02 &13.85\\
              &      &&14.09 &14.92\\
              &      &&16.49 &16.87\\
              &      &&17.37 &17.35\\
              &      &&23.36 &21.19\\
              &      &&37.75 &28.75\\
    \end{tabular}\label{tab:Cn}
\end{ruledtabular}
\end{center}
\end{table}

\subsection{Magnetization and magnetic charge} \label{sec:MandZm}

In the A$_2$ phase, the trimerization induces not only an electric
polarization, but also a weak ferromagnetism in the $\hat{z}$ direction
arising from out-of-plane tilting of the Mn$^{3+}$ spin moments induced
by SOC.  The net magnetizations in the 30-atom unit cell for
A$_2$-phase \HoMO, \ErMO, \YbMO, and \LuMO\ are
0.309, 0.303, 0.292, and 0.268\,$\mu_B$, respectively.
These magnetic moments are found to depend almost linearly on the
tilting angle of the MnO$_5$ bipyramids, which takes values of
5.03$^\circ$, 5.07$^\circ$, 5.16$^\circ$, and 5.21$^\circ$
respectively in these four compounds, but in any case the variation
is not very large. In contrast, the result for \LuFO\ is
-0.077\,$\mu_B$, which is much smaller and of opposite sign
compared with the \RMO\ materials.

The magnetic charges defined in Eq.~(\ref{eq:Zm}) are more sensitive
to the local environment, and now the differences between \RMO\
compounds are more significant.  We divide the magnetic charge
components into two groups that we label as ``longitudinal'' and
``transverse'' depending on whether the coupling is to magnetic
fields along the $\hat{z}$ direction or in the $x$-$y$ plane
respectively.\footnote{Note that this differs from the usual
convention for the magnetic susceptibility, where the distinction
between ``longitudinal'' and ``transverse'' corresponds to the
direction of the applied field relative to the spin direction.}

The longitudinal magnetic charge components are calculated with a
magnetic field directed along $\hat{z}$, which is roughly perpendicular
to the spin directions. These components are only non-zero when SOC is
considered.  The scenario here is similar to the case of a transverse
magnetic field ($H_x$ or $H_y$) applied to \CrO, since the magnetization
is along the $z$ axis for \CrO. It is therefore not surprising to find
that the longitudinal magnetic charges of \RMO\ and \LuFO\ in
Table~\ref{tab:Zmlongi} are comparable to the SOC-induced transverse
magnetic charges in \CrO. \cite{Ye14} The longitudinal magnetic charges
for \Op$_1$ and \Op$_2$ in \LuFO\ are opposite to, and about three times
smaller than, the ones in \RMO. These results explain the differences
between \RMO\ and \LuFO\ regarding the magnitude and the direction of
the weak ferromagnetism, which is generated by trimerization
distortions involving vertical displacements of \Op$_1$ and \Op$_2$.

\begin{table}
\begin{center}
\begin{ruledtabular}
    \caption{Longitudinal magnetic charge components $\Zm$
    ($10^{-3}\mu_{\rm B}/{\rm\AA}$) of \RMO\ and \LuFO\ in the A$_2$ phase.
    All components vanish in the absence of SOC.}
    \begin{tabular}{cddddd}
        &\multicolumn{1}{c}{\HoMO}
        &\multicolumn{1}{c}{\ErMO}
        &\multicolumn{1}{c}{\YbMO}
        &\multicolumn{1}{c}{\LuMO}
        &\multicolumn{1}{c}{\LuFO} \\ \hline
    $\Zm_{zz}$($R_1$)   &-50 &-53 &-53 &-67 & 7 \\
    $\Zm_{zz}$($R_2$)   & 14 & 35 & 24 & 16 & 7 \\
    $\Zm_{xz}$(TM)      &-92 &-86 &-61 &-67 & 9 \\
    $\Zm_{zz}$(TM)      & 24 &  1 &  6 & 25 & 2 \\
    $\Zm_{xz}$(\Ot$_1$) &-49 &-44 &-41 &-19 & 23\\
    $\Zm_{zz}$(\Ot$_1$) & 99 & 81 & 53 & 33 & 22\\
    $\Zm_{xz}$(\Ot$_2$) & -7 &-12 &-12 &-12 & 0 \\
    $\Zm_{zz}$(\Ot$_2$) &-119&-94 &-64 &-49 &-25\\
    $\Zm_{zz}$(\Op$_1$) &-276&-257&-230&-190& 54\\
    $\Zm_{zz}$(\Op$_2$) & 141& 140& 125& 100&-35\\
    \end{tabular}\label{tab:Zmlongi}
\end{ruledtabular}
\end{center}
\end{table}

\begin{table}
\begin{center}
\begin{ruledtabular}
    \caption{Transverse magnetic charge components $\Zm$
    ($10^{-2}\mu_{\rm B}/{\rm\AA}$) of \HoMO\ in the A$_1$
    phase, as computed inculding or excluding SOC.}
    \begin{tabular}{lddcldd}
        &\multicolumn{1}{c}{Total}
        &\multicolumn{1}{c}{No SOC} & &
        &\multicolumn{1}{c}{Total}
        &\multicolumn{1}{c}{No SOC}\\\hline
    $\Zm_{yx}$(Ho$_1$)  &-25&-28 && $\Zm_{zy}$(\Ot$_1$) &-188&-230\\
    $\Zm_{xx}$(Ho$_2$)  &-15&-18 && $\Zm_{yx}$(\Ot$_2$) &-57 &-67 \\
    $\Zm_{yx}$(Ho$_2$)  &-1 & 3  && $\Zm_{xy}$(\Ot$_2$) &-20 &-26 \\
    $\Zm_{yx}$(Mn)      & 92& 54 && $\Zm_{zy}$(\Ot$_2$) &-192&-231\\
    $\Zm_{xy}$(Mn)      &-10& 2  && $\Zm_{yx}$(\Op$_1$) &-483&-551\\
    $\Zm_{zy}$(Mn)      & 41& 48 && $\Zm_{xx}$(\Op$_2$) & 395& 461\\
    $\Zm_{yx}$(\Ot$_1$) & 23& 28 && $\Zm_{yx}$(\Op$_2$) & 184& 253\\
    $\Zm_{xy}$(\Ot$_1$) &-7 &-7  && \\
    \end{tabular}\label{tab:ZmtranA1phase}
\end{ruledtabular}
\end{center}
\end{table}

\begin{table*}
\begin{center}
\begin{ruledtabular}
    \caption{Transverse magnetic charge components $\Zm$
    ($10^{-2}\mu_{\rm B}/{\rm\AA}$) of \RMO\ and \LuFO\
    in the A$_2$ phase, as computed including or excluding SOC.}
    \begin{tabular}{ldddddddddd}
        &\multicolumn{2}{c}{\HoMO}&\multicolumn{2}{c}{\ErMO}&\multicolumn{2}{c}{\YbMO}&\multicolumn{2}{c}{\LuMO}&\multicolumn{2}{c}{\LuFO} \\
        &\multicolumn{1}{c}{Total} &\multicolumn{1}{c}{No SOC}
        &\multicolumn{1}{c}{Total} &\multicolumn{1}{c}{No SOC}
        &\multicolumn{1}{c}{Total} &\multicolumn{1}{c}{No SOC}
        &\multicolumn{1}{c}{Total} &\multicolumn{1}{c}{No SOC}
        &\multicolumn{1}{c}{Total} &\multicolumn{1}{c}{No SOC} \\ \hline
    $\Zm_{xx}$(R$_1$)   &-23 &-24 &-21 &-22 &-37 &-40 &-42 &-35 &-36 &-52 \\
    $\Zm_{xx}$($R_2$)   &6   &-1  &6   & 3  &12  & 9  &14  & 6  &15  & 24 \\
    $\Zm_{yx}$($R_2$)   &16  &18  &11  & 12 &10  & 10 &8   & 7  &-9  &-11 \\
    $\Zm_{xx}$(TM)      &-2  &10  &-7  &-10 &-16 &-21 &-11 & 1  &-52 &-43 \\
    $\Zm_{zx}$(TM)      &-42 &-24 &-38 &-22 &-25 &-34 &-31 &-17 &-102&-95 \\
    $\Zm_{yy}$(TM)      &-5  &46  &-7  & 32 &-22 & 27 &-32 & 15 &-16 &-11 \\
    $\Zm_{xx}$(\Ot$_1$) &5   &5   &6   & 6  &12  & 16 &14  & 11 & 0  & 0  \\
    $\Zm_{zx}$(\Ot$_1$) &191 &221 &150 & 154&162 & 178&150 & 122&128 & 105\\
    $\Zm_{yy}$(\Ot$_1$) &24  &23  &22  & 22 &31  & 33 &34  & 25 &15  & 11 \\
    $\Zm_{xx}$(\Ot$_2$) &20  &23  &16  & 19 &19  & 22 &17  & 12 &25  & 20 \\
    $\Zm_{zx}$(\Ot$_2$) &195 &217 &140 & 161&173 & 189&166 & 134&130 & 110\\
    $\Zm_{yy}$(\Ot$_2$) &-59 &-61 &-48 &-46 &-57 &-60 &-57 &-45 &-41 &-42 \\
    $\Zm_{xx}$(\Op$_1$) &-445&-510&-392&-422&-532&-602&-564&-499&-665&-609\\
    $\Zm_{xx}$(\Op$_2$) & 241&234 &215 & 202&298 & 299&316 & 247&388 & 356\\
    $\Zm_{yx}$(\Op$_2$) &-378&-422&-335&-355&-466&-506&-498&-427&-673&-621\\
    \end{tabular}\label{tab:ZmtranA2phase}
\end{ruledtabular}
\end{center}
\end{table*}

For the response to transverse magnetic fields, both the field and the
spins lie in the basal plane, so the dynamical magnetic charges are
driven by both SOC and exchange striction. As the exchange-striction
strength can exceed that of the SOC by orders of magnitude in some
materials, it is worthwhile to understand the relative size of these
two effects in \RMO\ and \LuFO. In Tables~\ref{tab:ZmtranA1phase}
and \ref{tab:ZmtranA2phase}
we present the transverse magnetic charges
induced with and without SOC in the A$_1$ and A$_2$ phases. It is
obvious that the SOC contributions are an order of magnitude smaller
for many transverse components. Similarly, the magnetic charges induced by
exchange striction are about ten times larger than the SOC-driven
longitudinal ones in Table~\ref{tab:Zmlongi}.
However, the SOC is crucial for the Mn atoms and it
even reverses the signs of their transverse magnetic charges.

\subsection{Magnetoelectric effect} \label{sec:ME}
We calculate the spin-lattice MEC from Eq.~(\ref{eq:MEspin-lat}) using
our computed Born charges, force-constant matrices, and magnetic charges.
The spin-electronic contributions are calculated based on the
$\partial P/\partial H$ version of
Eq.~(\ref{eq:ME}) with the lattice degrees of freedom frozen.  We
further subdivide the ME tensor components into longitudinal and
transverse ones based on the direction of $\bf H$ relative to the
hexagonal axis as before, so that the longitudinal (transverse)
spin-lattice MEC is calculated using the longitudinal (transverse)
magnetic charge components.
The MEC tensor elements
allowed by symmetry are the longitudinal $\alpha_{zz}$ and
transverse $\alpha_{xx}=\alpha_{yy}$ ones in the A$_2$ phase, and
only the transverse $\alpha_{yx}=-\alpha_{xy}$ components
in the A$_1$ phase.

\begin{table}[b]
\begin{center}
\begin{ruledtabular}
    \caption{Computed MECs $\alpha_{zz}$
    (longitudinal) and $\alpha_{xx}$ and $\alpha_{yx}$ (transverse)
    for \RMO\ and \LuFO\ (ps/m).  Spin-lattice,
    spin-electronic, and total spin couplings are given as
    computed with and without SOC.}
    \begin{tabular}{ldddddd}
    &\multicolumn{2}{c}{Spin-latt.}
    &\multicolumn{2}{c}{Spin-elec.}
    &\multicolumn{2}{c}{Total spin} \\
    &\multicolumn{1}{c}{Total}&\multicolumn{1}{c}{No SOC}
    &\multicolumn{1}{c}{Total}&\multicolumn{1}{c}{No SOC}
    &\multicolumn{1}{c}{Total}&\multicolumn{1}{c}{No SOC}\\ \hline
    \multicolumn{7}{c}{$\alpha_{zz}$ in A$_2$ phase} \\
    \HoMO $\phantom{a}$
               & -0.27 & 0 & 0.06 & 0 & -0.21 & 0 \\
    \ErMO & -0.26 & 0 & 0.05 & 0 & -0.21 & 0 \\
    \YbMO & -0.25 & 0 & 0.06 & 0 & -0.19 & 0 \\
    \LuMO & -0.19 & 0 & 0.00 & 0 & -0.19 & 0 \\
    \LuFO &  0.26 & 0 & 0.00 & 0 &  0.26 & 0 \\ 
    \multicolumn{7}{c}{$\alpha_{xx}$ in A$_2$ phase} \\
    \HoMO &-0.99&5.12 &4.10 & 4.83&3.11 &9.95 \\
    \ErMO &-1.30&2.40 &2.56 & 3.72&1.26 &6.12 \\
    \YbMO &-2.52&1.20 &3.72 & 4.66&1.20 &5.86 \\
    \LuMO &-2.60&1.31 &3.82 & 3.50&1.22 &4.81 \\
    \LuFO &-2.20&-1.57&-0.79&-0.32&-2.99&-1.89\\ 
    \multicolumn{7}{c}{$\alpha_{yx}$ in A$_1$ phase} \\
    \HoMO & 9.55 & 4.88 & 5.24 & 5.35 & 14.79 & 10.23 \\
    \end{tabular}\label{tab:MEall}
\end{ruledtabular}
\end{center}
\end{table}

\begin{figure}[b]
  \includegraphics[width=7.5cm]{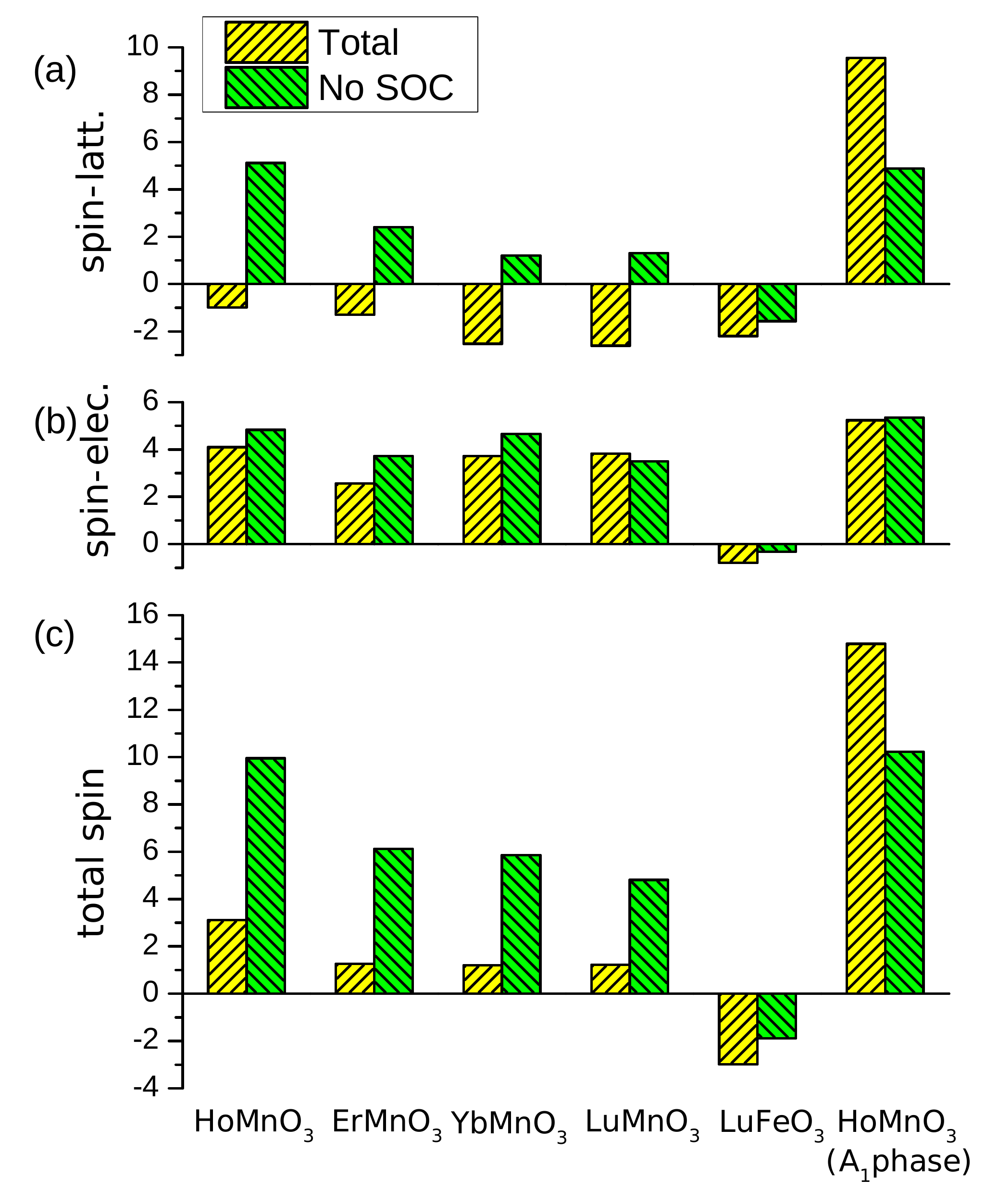}
    \caption{Transverse MECs for \RMO\ and \LuFO. $\alpha_{xx}$ (ps/m)
    in the A$_2$ phase and $\alpha_{yx}$ in the A$_1$ phase.
    (a) Spin-lattice; (b) spin-electronic; and (c) total spin couplings.}
  \label{fig:MEtransverse}
\end{figure}

In the first part of Table~\ref{tab:MEall}, the spin-contributed
longitudinal MECs are
shown for \RMO\ and \LuFO\ in the A$_2$ phase. The MEC from the spin
channel is dominated by the spin-lattice contribution. Although the
longitudinal magnetic charges of \LuFO\ are smaller than for \RMO, the
spin-lattice MECs $|\alpha_{zz}|$ in \RMO\ and \LuFO\ are similar,
$\sim$\,0.25\,ps/m. The results are comparable to those reported for
the transverse MEC in \CrO\cite{Malashevich12} and for
$\alpha_{zz}$ in \ErMO\ \cite{Das14} in previous first-principles
calculations.
In the second part of Table~\ref{tab:MEall}, we show the spin-related
transverse MECs $\alpha_{xx}$ for \RMO\ and \LuFO\ in the A$_2$ phase.
The same information is presented in graphical form in Fig.~\ref{fig:MEtransverse}.

It is clear from the comparison between the first and second parts of
Table~\ref{tab:MEall} that the transverse spin-lattice MECs are
one order of magnitude larger than the longitudinal ones, as a result
of the exchange-striction mechanism.
Surprisingly, Fig.~\ref{fig:MEtransverse}(a) shows that the effect of
SOC on the exchange striction is profound, even reversing the sign of
the spin-lattice MECs in \RMO.
This unusual behavior can be traced mainly to two observations about the
spin-lattice contributions from different IR-active modes in the \RMO\
materials.
Firstly, the exchange-striction MEC is smaller than expected as a
result of a large degree of cancellation between the contributions from
different transverse IR-active modes.
To illustrate this, the mode-by-mode contributions are
presented for a few selected cases in Table \ref{tab:MEmodes}.
Secondly, the softest modes are dominated by Mn displacements,
precisely those for which SOC has the largest effect on the $\Zm$ values,
even flipping the sign of some components.
In this way, it turns out that SOC can result in large relative changes
in the MEC results.
In the case of \LuFO, the SOC effect on the $\Zm$ values is weak, even
for Fe atoms. Thus, the MEC of \LuFO\ does not change as dramatically as
that of \RMO\ when SOC is included.

\begin{table}
\begin{center}
\begin{ruledtabular}
    \caption{Transverse MEC contributions (ps/m) from IR-active modes
    for A$_2$ and A$_1$ phases of \HoMO\ and A$_2$ phase of \LuMO.
    Results are given in ascending order of force-constant eigenvalues,
    which are reported in Table II of the Supplement.}
    \begin{tabular}{ddcddcdd}
        \multicolumn{2}{c}{A$_2$ phase \HoMO} &&\multicolumn{2}{c}{A$_1$ phase \HoMO} &&\multicolumn{2}{c}{A$_2$ phase \LuFO} \\
        \multicolumn{1}{c}{Total}&\multicolumn{1}{c}{No SOC}
        &&\multicolumn{1}{c}{Total}&\multicolumn{1}{c}{No SOC}
        &&\multicolumn{1}{c}{Total}&\multicolumn{1}{c}{No SOC}\\ \hline
         0.01& 0.12&& 0.25& 0.18&& 0.28& 0.39\\
        -1.16& 2.62&& 4.98& 2.36&&-0.54&-0.50\\
         0.66& 2.32&& 3.59& 2.37&&-1.31&-1.22\\
        -0.51&-0.35&&-0.32&-0.48&& 1.30& 1.23\\
         2.79& 3.13&& 2.87& 3.33&& 3.31& 3.12\\
         0.35& 0.21&& 0.30& 0.30&& 1.84& 1.73\\
        -1.88&-1.85&&-1.35&-1.90&&-4.43&-4.11\\
         1.13& 1.25&& 1.19& 1.38&&-2.59&-2.25\\
        -2.96&-3.07&&-2.70&-3.40&& 1.24& 1.13\\
         0.01& 0.13&& 0.19& 0.06&&-1.48&-1.27\\
         0.21& 0.24&& 0.21& 0.26&&-0.15&-0.14\\
         0.36& 0.40&& 0.34& 0.42&& 0.89& 0.83\\
        -0.03&-0.03&&-0.03&-0.04&&-0.62&-0.55\\
         0.02& 0.01&& 0.03& 0.03&& 0.07& 0.03
    \end{tabular}\label{tab:MEmodes}
\end{ruledtabular}
\end{center}
\end{table}

From Fig.~\ref{fig:MEtransverse}(b) it can be seen
that the spin-electronic contribution is not negligible in the transverse
direction, and it counteracts the MEC from the spin-lattice channel in
A$_2$ phase \RMO. The total transverse ME effect is summarized in
Fig.~\ref{fig:MEtransverse}(c).
Because of the large SOC effect and the cancellation between the
lattice and electronic contributions, the total spin MEC $\alpha_{xx}$
is $\sim$\,1.2\,ps/m in most A$_2$-phase \RMO\ compounds, except for
\HoMO. In \HoMO, the cancellation between the spin-lattice and the
spin-electronic MECs is the weakest of all the \RMO\ compounds,
resulting in the largest total spin MEC of $\sim$\,$3.1$\,ps/m in the
A$_2$ phase. In \LuFO, the spin-lattice and spin-electronic terms are
all smaller than in \RMO. However, the
cancellation induced by the SOC perturbation and the spin-electronic
contribution is avoided, so that \LuFO\ has a large total spin
MEC of $\sim$\,$-3$\,ps/m.

%
We present the MECs for \HoMO\ in the A$_1$ phase in the last line of
Table~\ref{tab:MEall} and in Fig.~\ref{fig:MEtransverse}. In principle
the MECs of \HoMO\ in the A$_1$ and A$_2$ phases should be the same
without SOC, as the two phases only differ by a global spin rotation.
This is approximately confirmed by a comparison of the corresponding
entries for \HoMO\ in Table\,\ref{tab:MEall}. The ME contribution from
exchange striction (i.e., without SOC) is $\sim$\,$5$\,ps/m for both
the A$_2$ and A$_1$ phases. However, when the effect of SOC is
included, the spin-lattice contribution is strongly enhanced by another
$\sim$\,$5$\,ps/m. Furthermore, the spin-electronic MEC has the same
sign as the spin-lattice one, which adds $\sim$\,$5$\,ps/m to the MEC.
Therefore, the total spin MEC $\alpha_{yx}$ reaches $\sim$\,$15$\,ps/m,
and is the largest in all of the \RMO\ and \LuMO\ materials we studied.

\section{Summary} \label{sec:summary}
In summary, we have studied the spin-related magnetic charges
and MECs for \HoMO, \ErMO, \YbMO, \LuMO, and \LuFO\ using
first-principles calculations. We confirm that the exchange
striction acting on noncollinear spins induces much larger
magnetic charges
than does SOC acting alone.  Nevertheless, the effect of
SOC on the MECs is surprisingly large, rivaling that of
exchange striction in many cases.  This occurs because
the exchange-striction contribution tends to be reduced by
cancellations between different IR-active modes, while the SOC
contribution is mainly associated with just a few low-frequency
modes with large Mn displacements.
We also find that the \RMO\ materials have spin-electronic MECs
comparable to the spin-lattice ones. Among the \RMO\ and \LuFO\
materials we studied, we find that the A$_1$ phase of \HoMO\ is the
most promising ME material, with the largest MEC of $\sim$\,15\,ps/m.
Extrapolating our conclusions to other hexagonal \RMO\ and \RFO\
compounds that are not included in our calculations,
we predict that the A$_2$ phase is more promising for the ferrites,
while the A$_1$ phase has a stronger MEC for the manganites.

\acknowledgments

We thank Weida Wu for useful discussions.
The work was supported by ONR grant N00014-12-1-1035.



\end{document}


\title{Supplementary material for ``Magnetic charge and the
magnetoelectricity in hexagonal manganites \RMO\ and ferrites \RFO''}

\author{Meng Ye}
\affiliation{Department of Physics \& Astronomy, Rutgers University,
Piscataway, New Jersey 08854, USA}

\author{David Vanderbilt}
\affiliation{Department of Physics \& Astronomy, Rutgers University,
Piscataway, New Jersey 08854, USA}


\begin{abstract}
In Tables I-II of the main text, we provided detailed information
on the Born charge tensors and force constant eigenvalues only
for the two representative materials \LuMO\ and \LuFO.  Here,
we provide the same information for the other materials covered
by this study.  Note that the values given in the last two columns of
each table are redundant with those given in Tables I-II of the
main text.
\end{abstract}

\maketitle

\begin{table}[h!]
\begin{center}
\begin{ruledtabular}
    \caption{Atomic Born charge tensors $\Ze$
    (in units of $|e|$) for \RMO\
    and \LuFO\ in the A$_2$ phase. TM\,=\,Mn, Fe.}
    \begin{tabular}{cddddd}
        &\multicolumn{1}{c}{\HoMO}
        &\multicolumn{1}{c}{\ErMO}
        &\multicolumn{1}{c}{\YbMO}
        &\multicolumn{1}{c}{\LuMO}
        &\multicolumn{1}{c}{\LuFO}\\ \hline
    $\Ze_{xx}$(R$_1$)   & 3.69& 3.67& 3.62& 3.61& 3.79\\
    $\Ze_{zz}$(R$_1$)   & 4.16& 4.15& 4.11& 4.12& 3.94\\
    $\Ze_{xx}$(R$_2$)   & 3.76& 3.73& 3.67& 3.66& 3.84\\
    $\Ze_{yx}$(R$_2$)   & 0.13& 0.13& 0.13& 0.13& 0.15\\
    $\Ze_{zz}$(R$_2$)   & 4.07& 4.05& 4.00& 3.96& 3.88\\
    $\Ze_{xx}$(TM)      & 3.16& 3.17& 3.17& 3.17& 2.96\\
    $\Ze_{zx}$(TM)      & 0.41& 0.42& 0.43& 0.44& 0.21\\
    $\Ze_{yy}$(TM)      & 3.25& 3.25& 3.26& 3.26& 3.01\\
    $\Ze_{xz}$(TM)      & 0.07& 0.07& 0.07& 0.07&-0.02\\
    $\Ze_{zz}$(TM)      & 4.02& 4.01& 3.97& 3.95& 4.16\\
    $\Ze_{xx}$(\Ot$_1$) &-1.95&-1.94&-1.92&-1.92&-2.19\\
    $\Ze_{zx}$(\Ot$_1$) & 0.24& 0.24& 0.24& 0.25& 0.25\\
    $\Ze_{yy}$(\Ot$_1$) &-2.05&-2.03&-2.00&-2.00&-2.28\\
    $\Ze_{xz}$(\Ot$_1$) & 0.19& 0.19& 0.19& 0.19& 0.11\\
    $\Ze_{zz}$(\Ot$_1$) &-3.24&-3.24&-3.20&-3.19&-3.21\\
    $\Ze_{xx}$(\Ot$_2$) &-1.95&-1.93&-1.91&-1.90&-2.15\\
    $\Ze_{zx}$(\Ot$_2$) &-0.20&-0.20&-0.20&-0.20&-0.19\\
    $\Ze_{yy}$(\Ot$_2$) &-1.88&-1.87&-1.85&-1.85&-2.13\\
    $\Ze_{xz}$(\Ot$_2$) &-0.18&-0.18&-0.18&-0.18&-0.11\\
    $\Ze_{zz}$(\Ot$_2$) &-3.38&-3.38&-3.34&-3.33&-3.30\\
    $\Ze_{xx}$(\Op$_1$) &-3.01&-3.01&-3.01&-3.00&-2.40\\
    $\Ze_{zz}$(\Op$_1$) &-1.58&-1.57&-1.54&-1.54&-1.61\\
    $\Ze_{xx}$(\Op$_2$) &-3.05&-3.05&-3.06&-3.05&-2.45\\
    $\Ze_{yx}$(\Op$_2$) &-0.03&-0.03&-0.03&-0.03&-0.02\\
    $\Ze_{zz}$(\Op$_2$) &-1.47&-1.46&-1.43&-1.43&-1.52\\
    \end{tabular} \end{ruledtabular}
\end{center}
\end{table}

\newpage

\begin{table}[h!]
\begin{center}
\begin{ruledtabular}
    \caption{Eigenvalues of the force-constants matrix  ($\rm{eV}/{\AA}^2$)
    for IR-active modes in \RMO\ and \LuFO\ in the A$_2$ phase, and for
    \HoMO\ in the A$_1$ phase}
    \begin{tabular}{ddddd}
        \multicolumn{1}{c}{\HoMO} &
        \multicolumn{1}{c}{\ErMO} &
        \multicolumn{1}{c}{\YbMO} &
        \multicolumn{1}{c}{\LuMO} &
        \multicolumn{1}{c}{\LuFO} \\ \hline
        \multicolumn{5}{l}{Longitudinal $A_1$ modes}\\
         4.23 &  4.23 & 4.25 &4.24  &3.48  \\ 
         7.11 &  7.18 & 7.35 &7.44  &6.70  \\ 
         8.14 &  8.27 & 8.60 &8.74  &8.41  \\ 
        10.77 & 10.90 &11.34 &11.51 &11.47 \\
        13.69 & 13.82 &13.98 &14.01 &12.03 \\
        14.85 & 15.03 &15.42 &15.60 &15.59 \\
        21.32 & 21.60 &22.36 &22.66 &20.53 \\
        25.44 & 25.57 &25.67 &25.87 &22.83 \\
        35.99 & 35.68 &35.54 &35.82 &28.46 \\
      \multicolumn{5}{l}{Transverse $E_1$ modes}\\
         3.23 & 3.37 & 3.27 & 3.32 & 3.56 \\
         4.22 & 4.25 & 4.49 & 4.68 & 4.62 \\
         5.96 & 6.28 & 6.63 & 6.73 & 6.97 \\
         7.59 & 6.93 & 7.01 & 7.35 & 8.09 \\         
         8.41 & 8.56 & 8.57 & 8.63 & 8.83 \\
         9.29 & 8.99 & 9.31 & 9.56 & 9.24 \\
         9.65 &10.12 &10.95 &11.36 &11.37 \\
        11.23 &11.25 &12.02 &12.46 &12.46 \\
        12.57 &12.85 &12.95 &13.02 &13.85 \\
        13.29 &13.54 &13.77 &14.09 &14.92 \\
        16.41 &16.76 &16.57 &16.49 &16.87 \\
        17.49 &17.52 &17.38 &17.37 &17.35 \\
        22.79 &23.02 &23.16 &23.36 &21.19 \\
        36.18 &37.99 &37.54 &37.75 &28.75 \\
    \end{tabular} \end{ruledtabular}
\end{center}
\end{table}